# Prussian Blue Modified Reduced Graphene Oxide as Support for Pt Nanoparticles: Development of Efficient Catalysts for Oxygen Electroreduction in Acid Medium.

B. Zakrzewska[a], B. Dembinska[a], S. Zoladek[a], I. Rutkowska[a], J. Żak[a], L. Stobinski[a,b], A. Małolepszy[b], E. Negro[c], V. Di Noto[c], P.J. Kulesza[a], K. Miecznikowski[a*]

[a]Faculty of Chemistry, University of Warsaw, Pasteura 1, 02-093 Warsaw, Poland
[b]Faculty of Chemical and Process Engineering, Warsaw University of Technology, Warynskiego 1, 00-645 Warsaw, Poland
[c]Department of Industrial Engineering, Università degli Studi di Padova in Department of Chemical Sciences, Via Marzolo 1, 35131 Padova (PD), Italy

___

*Corresponding Author. Tel.: +48-22-5526340; Fax: +49-22-8225996; E-mail address: kmiecz@chem.uw.edu.pl

**Abstract**

Pt electrocatalytic nanoparticles were deposited onto hybrid carriers composed of reduced graphene oxide (rGO) – transition metal hexacyanoferrate (Prussian Blue – PB) and the resulting system's electrochemical activity was investigated during oxygen reduction reaction in acidic solution. The Prussian Blue -utilizing and Pt nanoparticle-containing materials were characterized using transmission electron microscopy, X-ray diffraction and electrochemical diagnostic techniques such as cyclic voltammetry and rotating ring-disk voltammetry. Application of rGO carriers modified with Prussian Blue as matrices (supports) for Pt catalytic centers does not change practically the potential of electroreduction of oxygen in 0.5 mol dm$^{-3}$ H$_2$SO$_4$ (under rotating disk voltammetric conditions) relative to the behavior of the analogous PB-free system. What is even more important that, due to the presence of the polynuclear cyanoferrate modifier, the amounts of the undesirable hydrogen peroxide intermediate are significantly decreased (at ring in the rotating ring-disk voltammetry). The results are consistent with the bifunctional mechanism in which oxygen reduction is initiated at Pt centers and the hydrogen peroxide intermediate is reductively decomposed at reactive PB-modified rGO supports.

**Introduction**

Proton Exchange Fuel Cells have been viewed as suitable power sources candidates for stationary and vehicle propulsion due to their high efficiency, high energy density and low temperature energy conversion. The reactions in fuel cells invariably involve oxygen reduction at the cathode side. It should be note that oxygen reduction reaction, in the context of energy, is the challenging reaction due to their rate limiting steps in the process of getting energy from fuel. The sluggish oxygen reduction has still a hugely affect on losing efficiency, the power output and lifetime of the fuel cells [1,2]. Pt-based catalysts, mostly applied and studied for their high efficiency with respect to other Pt-free catalysts in acid medium consist of Pt nanoparticles (NPs) supported onto different supports,.e.g., carbon, oxides and carbon/oxides composites. While great progress has been made, there are still issues to be concerned about. For instance: (i) low resistance of the Pt catalyst to poisoning (mainly to CO), (ii) hydrogen peroxide production at the cathode that degrades fuel cell membranes and (iii) methanol crossover that results in mixed potential.

The use of a suitable supporting material enables a high dispersion of metal nanoparticles onto its surface, which is particularly important with expensive noble metal catalysts, such as Pt and Pd. The appropriate supporting materials should be characterized by the following features: high surface area, corrosion resistance, and a stable surface to prevent the nanoparticles' agglomeration. A number of various supports have been extensively studied as high oxidation resistant materials, namely, graphitized carbon [3], conductive polymers [4] as well as conductive metal oxides such as ($MnO_x$ [5], $TiO_2$ [6-8], $CeO_x$ [9], ZnO [10], $WO_3$ [11-13], $SnO_2$ [14] and $NbO_2$ [15]). Carbon black (Vulcan XC-72) has been commonly used as a conventional supporting material for the Pt-based catalysts for fuel cell applications, due to its high surface area and conductivity, along with its low cost [16]. One drawback of the Vulcan support is that it is not very tolerant of fuel cell conditions, and is electrochemically oxidized to its surface oxides. In consequence, the oxidation process leads to the formation of large particles, resulting in a decrease in the fuel cell's performance [17]. To avoid these problems, various kind of graphene materials have been introduced to replace carbon black (Vulcan) as supporting materials in electrocatalysis, due to their increased tolerance for carbon corrosion, and their excellent mechanical, electronic, and surface properties [18, 19].

Such new catalysts also have to characterized by a minimum formation of $H_2O_2$ because free radicals that are produced from the breakdown of $H_2O_2$ have a detrimental effect on the polymer membranes – Nafions in fuel cell [20,21]. One of the methods to reduce the amount of hydrogen peroxide is to use Prussian Blue as the catalyst cathode additive [22,23]. In the literature has been many confirmation that PB has been an excellent electrocatalyst for hydrogen peroxide compared with the most conventional noble metals (e.g. Pt) or metal oxide materials [24-27]. Such properties of PB is connected with its analogy with the biological group of peroxidase, that are natural enzymes for the reduction of hydrogen peroxide, it has been called an artificial peroxidase [25]. Moreover, PB systems detect hydrogen peroxide at low applied potential (approx. 0.2 vs. RHE) what allow to enhance the immunity to interference (e.g. ascorbic acid, uric acid) as well as reduce its in the presence of oxygen, that is advantages in the contrast for Pt [24]. The kinetics of hydrogen peroxide reduction on PB-modified electrodes was investigated using a rotating disk electrode and in situ Raman spectroelectrochemical technique [26,27]. The Raman measurement showed that during reduction of hydrogen peroxide PB form exists in the layer even at potentials corresponding to the completely reduced form of PB and the process corresponded first-order reaction. Whereas, the result obtained from rotating disk electrode methodology exhibited two step reaction mechanism. The authors claim that first step is related with dissociative adsorption of hydrogen peroxide, where OH radicals are formed and the second step, the OH radicals are reduced to hydroxyl ions (one-electron process). The late step gets rate limiting factor in the presence of high concentration of peroxide and high pH.

In the present work, we explore the ability of Prussian Blue structures to effectively decompose, i.e. reduce readily at fairly high potentials (in acid medium) hydrogen peroxide that typically appears as undesirable intermediate during reduction of oxygen in fuel cells. When Prussian Blue structures are combined with a catalyst of moderate activity, such as rGO supported Pt nanoparticles, a highly reactive composite system for the oxygen reduction is produced [28]. It is reasonable to expect that rGO modified with Prussian Blue and supported Pt acts as a bifunctional system, in which oxygen reduction is primarily induced at Pt catalytic centers, whereas reduction any hydrogen peroxide intermediate is further activated at the Prussian Blue matrix. The data obtained are discussed in relation to the beneficial effect of Prussian Blue on oxygen reduction reaction. Electrochemical testing of catalysts has been carried out using a rotating ring disk electrode (RRDE) to show direct measurements of $H_2O_2$ production and yield an accurate assessment of the catalytic activity for ORR. Further, comparison has been made to the behavior of rGO-supported Pt nanoparticles under the analogous RRDE voltammetric conditions.

**Experimental**

All chemicals were commercial materials of the highest available purity and were used as received. In this study, a commercial graphite powder (ACROS ORGANICS) was used. Graphene oxide (GO) was prepared using modified Hummers method. In this process, 10 g of graphite powder was added to 230 mL of concentrated sulfuric acid (98 wt.%) and stirred for

30 minutes. Next 4.7 g of sodium nitrate and 30 g of potassium permanganate were slowly added to the mixture and the temperature was kept below 10ºC in an ice bath. Then the mixture was slowly heated up to ~33ºC and was controlled so as not to exceed 35ºC for 2 hours under stirring. In the next step, 100 mL of water was added to the mixture and the temperature reached ~120ºC. Finally, the mixture was treated with 10 mL of $H_2O_2$ (30 wt.%). Obtained slurry was kept in an ultrasonic bath for 1 hour. For purification, the slurry was filtered using ceramic membranes with 0,2 micron pore size and washed with deionized water in order to remove the by-products of the synthesis till the pH of the filtrate reached 6.5. Reduced graphene oxide (rGO) was obtained by adding 10 mL of 50% hydrazine water solution to 100 mL of 0,5 wt.% GO water dispersion. The mixture was heated up to 100ºC and kept under stirring for 2h. After reduction, the product was filtered using polyethersulfone (PES) filter with 0,8 μm pore size. Solutions were prepared from triply-distilled subsequently-deionized water. They were deaerated (using argon) or saturated with oxygen for at least 20 min prior to the electrochemical experiment. Experiments were conducted at room temperature (20 ± 0.5°C).

For the preparation of reduced graphene oxide (rGO) modified Prussian Blue and supported Pt (rGOPB/Pt) materials. The hybrid material (rGO modified PB) was produced as follows. A known amount of rGO (200 mg) was dispersed in 6 cm$^3$ solution of Prussian Blue (that had been obtained by mixing a solution 2 mmol dm$^{-3}$ $K_3[Fe(CN)_6]$, 2 mmol dm$^{-3}$ $FeCl_3$, and 50 mmol dm$^{-3}$ EDTA). Formation of PB nanostructure on rGO occurred through the sol-gel aggregation aging process. After 24 h mixing the following suspension was centrifuged and dried at 300ºC for 2 h. The obtained support hybrid material (PB-modified rGO) was used to fabricated Pt nanoparticles.

In order to obtained platinum nanoparticles deposited on the surface of the rGO reduction or rGO/PB methods were used. For this purpose the appropriate amount of the platinum potassium hexachloroplatinate ($K_2PtCl_6$) salt was added to the 0.5g rGO or rGO/PB water dispersion under continuous magnetic stirring. After 15 min the 0.1M KOH was added to the mixture to increase the pH to 10. Nanoparticles of Pt were obtained by adding 10 mL of 1M sodium borohydride (NaBH4) water solution to the slurry and kept under stirring. After 1h the 20 wt.% Pt/rGO catalyst deposit was filtered using PES filter with 0,8 μm pore size and washed with deionized water. Their composition of the final material was as follows: Pt 20% and rGO/PB 80% (by mass). The diameters of PB-modified rGO supports Pt nanoparticles were approximately 20 and 2 nm, respectively.

The electrochemical experiments were performed with CH Instruments (Austin, TX, USA) Model 750A workstation. A mercury/mercury sulfate electrode ($Hg/Hg_2SO_4$), which potential was 700 mV more positive relative to the Reversible Hydrogen Electrode (RHE), was used as a reference electrode. All potentials are expressed against the reversible hydrogen electrode (RHE).

Rotating disk electrode (RDE) and rotating ring disk electrode (RRDE) voltammetric measurements were accomplished using a variable speed rotator (Pine Instruments, USA). The electrode assembly utilized a glassy carbon disk and a Pt ring. The electrodes were polished with successively finer grade aqueous alumina slurries (grain size, 5-0.05 μm) on a Buehler polishing cloth. In the RRDE measurements, the radius of the disk electrode was 2.5 mm, and the inner and outer radii of the ring electrode were 3.25 and 3.75 mm, respectively. The collection efficiency of the RRDE assembly was determined from the ratio of ring and disk currents (at various rotation rates) using the argon-saturated 0.005 mol dm$^{-3}$ $K_3[Fe(CN)_6]$ + 0.01 mol dm$^{-3}$ $K_2SO_4$ solution

[29-33]. Based on five independent experiments, it was found that, within the potential range considered here, and at rotation rates up to 2500 rpm, the experimental collection efficiency (N) remained unchanged and was equal to 0.39. During the RRDE measurements in oxygen saturated solutions, the potential of the ring electrode was kept at 1.2 V (vs. RHE). At this potential, the generated $H_2O_2$ is oxidized under diffusional control. All RRDE polarization curves were recorded at a scan rate of 10 mV s$^{-1}$. Current densities were calculated with respect to the geometric surface area of the electrode.

To produce an suspension of PB-modified rGO supported Pt nanoparticles, a known amount (5 mg) of the catalyst was dispersed in 1 ml of 2-propanol and Nafion (20% by weight) as solvent and binder, respectively. The following suspension was subjected to magnetic stirring for 24 h. As a rule, 2 µl aliquot of the appropriate suspension was introduced onto the surface of a glassy carbon disk electrode (geometric area, 0.2 cm$^2$), and the suspension was air-dried at room temperature. Typical loadings of supports were in the range 10 µg cm$^{-2}$. As a rule the catalytic films were activated by performing 25 full voltammetric potential cycles in the potential range from 0 to 1 V (at 50 mV s-1) until steady-state currents were observed. The systems were systematically characterized to obtain information on particle morphology, composition, and crystal structure by various techniques, such as High Resolution Transmission Electron Microscopy (HR-TEM), Energy Dispersion Spectroscopy (EDS), and X-ray diffraction (XRD). The HR-TEM measurements were performed using an instrument with an accelerating voltage of 200 keV with EDS analysis (Bruker Quantax 400, SEM-EDS). Samples for TEM measurements were prepared by depositing drops of diluted colloidal solutions of nanoparticles onto 400-mesh copper grids that supported a Formvar film (Agar Scientific); they were dried under ambient laboratory conditions (temperature, 22 ± 1 °C) for 24 h prior to TEM analysis. The crystallographic phase analysis was carried out by XRD using a Bruker D8 Discover system operated with a Cu X-ray tube (1.5406 Å) and Vantec (linear) detector (k = 1.5406 Å). The lattice parameter value and particle size were obtained from the position and the full-width at half-maximum (FWHM) of the (220) peak.

**Results and discussion**

The X-ray diffraction examination was performed to confirm the presence as well as to obtain the information about the crystalline structure of the proposed electrocatalysts. Figure 1 displays the profiles of rGO_PB supported Pt nanoparticles, Pt/rGO and rGO_PB. Material exhibited diffraction peaks in the range $2\theta = 24.5°$, which were attributed to carbon-supported material with a (111) reflection plane for all samples; the PB and Pt reflections also arose in this [34]. In the case of both Pt/rGO and Pt/rGO_PB (Fig.1a and c), diffraction peaks at the corresponding diffraction angles (39.8, 46.3, 67.6, and 81.4) occurred, which were attributed to the (111), (200), (220), and (311) planes, characteristic of the respective crystalline faces of Pt (JCPDS No. 01-087-0646) and were decorated uniformly on the rGO surface. While for the rGO_PB and Pt/rGO_PB nanoparticles (Fig.1b and c), the other diffraction peaks appeared at $2\theta = 38.5, 44.7, 65.1,$ and $78.4°$ (JCPDS No. 52-1907); these are associated with the (420), (422), (640), and (301) reflections respectively, thus validating the formation of Prussian Blue [35]. It can be observed that PB signals are narrow; this can be attributed to the good crystalline structure and due to the very broad size of the particles. To calculate the average particle size according to the Scherrer equation, the (220) reflections of Pt were used. The average Pt particle size obtained from XRD measurements for Pt/rGO and Pt/rGO_PB was of the order of 7-10 nm for both catalysts and is consistent with the results from HR-TEM experiments.

The morphologies of the rGO_PB, Pt/rGO and Pt/rGO_PB were examined by HR-TEM (Fig.2). Figure 2 illustrates the transparent and pleated paper like structure of rGO nanosheets before successful decoration of Pt nanoparticles and PB structure. Fig. 2 displays dark Pt spots deposited on the reduce graphene oxide layers were spherical metal nanoparticles of similar sizes. The size of metal particles roughly estimated from HR-TEM images of ca. 7 nm, which is in god agreement with the XRD results. The inter-planar spacing 0.3940 nm (200) is attributed to Pt, which agree with literature values.

Fig. 3 presents the cyclic voltammetric responses of reduced graphene oxide (rGO) supported Pt (curve a) and PB-modified rGO supported Pt nanoparticles (curve b) deposits on glassy carbon electrode recorded in the argon-saturated 0.5 mol dm$^{-3}$ H$_2$SO$_4$. While the cyclic voltammogram of PB-free catalyst is characterized by a fairly flat background with the broad oxide reduction peak at the potential range from 0.2 V to 0.6V response (Fig.3a), the response of PB-modified rGO supported Pt system (Fig.3b) has yielded fairly high voltammetric currents at the whole potential range with the similar broad cathodic peak. In the case of PB-modified rGO supported Pt nanoparticles the presence of Prussian Blue active centers on carbon materials can be postulated as evidenced from higher voltammetric current with the PB surface at potentials lower than 0.15 V vs. RHE in comparison to PB-free catalysts.

The representative RDE voltamograms recorded at different rotation rates (ranging from 400 to 2500 rpm) for the oxygen reduction reaction (ORR) at catalytic layers containing PB-modified rGO supported Pt nanoparticles are illustrated in Fig.4. As we can see, the RDE responses are fairly well-defined at any rotation rate studied but the mass-transport limited currents are not reached over the whole potential region explored. Similar electrochemical behavior for ORR was reported in literature for Pt/Vulcan catalysts [32]. Further, half-wave potentials for the reduction of oxygen at the system containing PB-modified rGO supported Pt nanoparticles are not shifted when compared to those characteristic of the PB-free system.

In order to extract kinetic information, the RDE curves were analyzed using the Levich and Koutecki-Levich plots (Fig.5). When the dependencies (Levich type plots – not show) of the RDE current densities (measured at 0.7 V) have been plotted versus square root of rotation rates for electrodes covered with PB-modified rGO supported Pt nanoparticles, the deviation from linearity (i.e. from the ideal behavior characteristic of systems limited solely by convective diffusion of oxygen in solution) is observed, and it seems to be more pronounced at bare rather than PB-modified system (not show here). Apparently, the kinetic control is more pronounced, i.e. the catalytic reaction is slower or less effective in the case of bare rGO/Pt nanoparticles. The reciprocal values of the experimental currents are plotted vs. $\omega^{-1/2}$ (Koutecky-Levich plots) for different electrode potentials at 0.7 V (Fig.5) [11,12,32]. Such diagnosis is justified because charge (electron, proton) propagation within the catalytic film should be fast, and the oxygen reactant is expected to have easy access to the dispersed active sites (bare or PB-modified rGO supported Pt nanoparticles). With low amounts of Nafion used to bind the catalyst to the substrate, any potential limitations related to mass transport through mixed PB-Nafion layers on the measured current densities are considered as relatively negligible [36,37]. We assume that only such factors as transport of oxygen in solution or, at higher rotation rates, the dynamics of the chemical (catalytic) step (reaction) are rate determining. The reciprocal plots (Fig.5) have yielded non-zero intercepts clearly indicating kinetic limitations associated with the electrocatalytic film.

RDE voltammetric limiting-current densities (jlim) can be expressed as follows [23,38]:

$$\frac{1}{j_{lim}} = \frac{1}{nFAkC_{film}C_{O_2}} + \frac{1}{j_L}$$

where $j_L$ is given by the Levich equation (in which the convective diffusion component is proportional to square root of rotation rate), k is the rate constant for the catalytic reaction in homogeneous units, $C_{film}$ is the surface concentration of the electrocatalyst, and $C_{Ox}$ is the bulk concentration of oxygen (1.1 mmol dm$^{-3}$ [39]). The symbols n and F stand for the number of electrons involved in the process (n = 4) and the Faraday constant. The kinetic parameters can be found from intercepts of the reciprocal plots (Fig.5). The estimated value of kCfilm, which is equivalent to the intrinsic rate of heterogeneous charge transfer, is $1 \times 10^{-1}$ cm s$^{-1}$ for the catalytic electroreduction (at 0.7 V) of oxygen at glassy carbon disks covered with PB-modified rGO supported Pt nanoparticles.

In the context of electrocatalysts for cathode for fuel cells an important issue is the degree of formation of the hydrogen peroxide intermediate during ORR in the presence and absence of Prussian Blue. In this case we have performed rotating ring-disk electrode (RDDE) measurements in which the catalytic nanoparticles were deposited onto disk electrodes. The responses recorded (upon application of the oxidative potential of 1.2 V vs RHE) at the ring Pt electrode are consistent with the view that formation of hydrogen peroxide intermediate have significantly decreased following PB-modified rGO supported Pt nanoparticles (Fig.6). This effect is particularly evident at potentials lower than 0.5 V where the Prussian Blue (PB) starts to be reduced to Prussian White that are capable of inducing reduction of $H_2O_2$ (to $H_2O$). The latter observation suggests direct involvement of Prussian Blue matrix in the decomposition of hydrogen peroxide intermediate and promoting effectively the almost 4-electron reduction of oxygen to water. A bifunctional mechanism, in which both Pt and PB are reactive, namely towards oxygen and hydrogen peroxide, respectively, is possible.

More quantitative information, namely in the % amounts of $H_2O_2$ ($X_{\%H2O2}$), about the relative degrees of formation of hydrogen peroxide during the oxygen reduction under the RRDE voltammetric conditions has been obtained using the equation described earlier [32,40]:

$$X_{\%H_2O_2} = \frac{2 \cdot \frac{I_r}{N}}{I_d + \frac{I_r}{N}} \cdot 100$$

where $I_r$ is ring current, $I_d$ stands for disk current, and N is the collection efficiency. The results are plotted versus potential applied to disk electrode in Fig.7. At the electrode modified with PB-modified rGO supported Pt nanoparticles (Curve b), the production of $H_2O_2$ is significantly lower in comparison to the same system without Prussian Blue (Curve a). At potentials lower than 0.55 V, the values of $X_{\%H2O2}$ are on the level 4 % for PB-modified rGO supported Pt nanoparticles. At potentials higher than 0.7 V, the oxygen reduction currents (at disk electrode) are not well-developed yet because the potentials are not sufficiently negative to drive effectively the oxygen reduction reaction; consequently, the values of $X_{\%H2O2}$ become somewhat higher. It should be remembered that the heterogeneous rate constant for the reduction of $H_2O_2$ at PB film on glassy carbon electrode is on the level $1 \times 10^{-1}$ cm s$^{-1}$. The latter value is comparable to that determined in the case of reduction of oxygen at Pt/rGO nanoparticles. This suggests that Prussian Blue matrix can be directly involved in the reductive decomposition of hydrogen peroxide intermediate and promoting effectively the almost 4-electron reduction of oxygen to water.

The overall number of electrons exchanged per O$_2$ molecule (n) was calculated as a function of the potential using the RRDE voltammetric data of Fig. 8 and using equation [41,42]:

$$n = 4I_{disk} / I_{disk} + I_{ring}/N$$

The corresponding number of transferred electron (n) per oxygen molecule involved in the oxygen reduction was estimated to be as follows: (A) 3.8–3.95, and (B) 3.9–3.95, for the oxygen reduction at Pt/rGO nanoparticles, and PB-modified rGO supported Pt nanoparticles (Fig. 8). The results demonstrate that, while the approximately four-electron reduction mechanism (with small formation of H$_2$O$_2$ intermediate) is the dominating pathway for the oxygen reduction (in acid medium) at all two types of Pt-containing catalytic systems, utilization of the chemically-reduced graphene-oxide based supports produces the hybrid system exhibiting the relatively highest activity toward electroreduction of oxygen in acid medium (Fig. 8).

**Conclusions**

On the basis of our RRDE diagnostic experiments it can be concluded that modification of rGO (support material) with ultra-thin layers of Prussian Blue results in the activation of Pt towards oxygen reduction. The phenomenon can originate from the properties of PB and its ability to catalyze reduction of hydrogen peroxide intermediate. The production of hydrogen peroxide has been found to be relatively low as measured by RRDE. Specific interactions between PB and support material (reduced graphene oxide - rGO), though not documented as yet, cannot be either excluded. Further research is along this line.

**Acknowledgements**

This work was supported by the European Commission through the Graphene Flagship – Core1 Project (GA-696656). The Polish side appreciates support from National Science Center (Poland) under Project No. 2015/19/B/ST4/03758

**References**

1. G. Hoogers, Fuel cell technology handbook, CRC Press, 2003.

2. A.F. Bruijn, G.J.M. Janssen, PEM fuel cell materials: cost, performance and durability, in: K.D. Kreuer (Ed.), Fuel Cells, Springer, New York, 2012, pp. 249-303.

3. E. Antolini, Formation of carbon-supported PtM alloys for low temperature fuel cells: a review, Mater. Chem. Phys. 78 (2003) 563-573.

4. B. Rajesh, K. Ravindranathan Thampi, J.M. Bonard, H.J. Mathieu, N. Xanthopoulos, B. Viswanathan, Electronically conducting hybrid material as high performance catalyst support for electrocatalytic application, J. Power Sources 141 (2005) 35-38.

5. N.R. Elezovic, B.M. Babic, V.R. Radmilovic, L.M. Vracar, N.V. Krstajic, Synthesis and characterization of $MoO_x$-Pt/C and $TiO_x$-Pt/C nano-catalysts for oxygen reduction Electrochim. Acta 54 (2009) 2404-2409.

6. L. Timperman, Y.J. Feng, W. Vogel, N. Alonso-Vante, Substrate effect on oxygen reduction electrocatalysis, Electrochim. Acta 55 (2010) 7558-7563.

7. N.R. De Tacconi, C.R. Chenthamarakshan, K. Rajeshwar, W.Y. Lin, T.F. Carlson, L. Nikiel, W.A. Wampler, S. Sambandam, V. Ramani, Photocatalytically Generated Pt/C–$TiO2$ Electrocatalysts with Enhanced Catalyst Dispersion for Improved Membrane Durability in Polymer Electrolyte Fuel Cells, J. Electrochem. Soc. 155 (2008) B1102-B1109.

8. W. Vogel, L. Timperman, N. Alonso-Vante, Probing metal substrate interaction of Pt nanoparticles: Structural XRD analysis and oxygen reduction reaction, Applied Catalysis A 377 (2010) 167-173.

9. Y. Luo, L. Calvillo, C. Daiguebonne, M.K. Daletou, G. Granozzi, N. Alonso-Vante, A highly efficient and stable oxygen reduction reaction on Pt/$CeO_x$/C electrocatalyst obtained via a sacrificial precursor based on a metal-organic framework, Appl. Catal. B: Environmental 189 (2016) 39–50.

10. B. Ruiz Camacho C. Morais, M.A. Valenzuela, N. Alonso-Vante, Enhancing oxygen reduction reaction activity and stability of platinum via oxide-carbon composites, Catal. Today 202 (2013) 36– 43.

11. P. Kulesza, K. Miecznikowski, B. Baranowska, M. Skunik, A. Kolary-Zurowska, A. Lewera, K. Karnicka, M. Chojak, I. Rutkowska, S. Fiechter, P. Bogdanoff, I. Dorbandt, G. Zehl, R. Hiesgen, E. Dirk, K. Nagabhushana, H. Boennemann, Electroreduction of oxygen at tungsten oxide modified carbon-supported $RuSe_x$ nanoparticles, J. Appl. Electrochem. 37 (2007) 1439-1446.

12. A. Lewera, K. Miecznikowski, R. Hunger, A. Kolary-Zurowska, A. Wieckowski, P.J. Kulesza, Electronic-level interactions of tungsten oxide with unsupported Se/Ru electrocatalytic nanoparticles, Electrochim. Acta 55 (2010) 7603–7609.

13. A. Lewera, L. Timperman, A. Roguska, N. Alonso-Vante, Metal-Support Interactions between Nanosized Pt and Metal Oxides ($WO_3$ and $TiO_2$) Studied Using X-ray Photoelectron Spectroscopy, J. Phys. Chem. C 115 (2011) 20153–20159.

14. B. Seger, A. Kongkanand, K. Vinodgopal, P.V. Kamat, Platinum dispersed on silica nanoparticle as electrocatalyst for PEM fuel cell, J. Electroanal. Chem. 621 (2008) 198–204.

15. K. Sasaki, L. Zhang, R.R. Adzic, Niobium oxide-supported platinum ultra-low amount electrocatalysts for oxygen reduction, Phys. Chem. Chem. Phys. 10 (2008) 159.

16. E. Antolini, Carbon supports for low-temperature fuel cell catalysts, Appl. Catal. B 88, (2009) 1-24.

17. K.H. Kangasniemi, D.A. Condit, T.D. Jarvi, Characterization of Vulcan Electrochemically Oxidized under Simulated PEM Fuel Cell Conditions, J. Electrochem. Soc. 151 (2004) E125-E132.

18. S. Litster, G. McLean, PEM fuel cell electrodes, J. Power Sources 130 (2004) 61-76.

19. D.A. Stevens, J.R. Dahn, Electrochemical Characterization of the Active Surface in Carbon-Supported Platinum Electrocatalysts for PEM Fuel Cells, J. Electrochem. Soc. 150 (2003) A770-A775.

20. D.A. Schiraldi, Perfluorinated Polymer Electrolyte Membrane Durability, Polymer Reviews 46 (2006) 315-327

21. S.J. Hamrock, M.A. Yandrasits, Proton Exchange Membranes for Fuel Cell Applications, Polymer Reviews 46 (2006) 219-244

22. P.J. Kulesza, L.R. Faulkner, Electrocatalytic Properties of Bifunctional Pt/W(VI,V) Oxide Microstructures Electrodeposited on Carbon Substrates, J. Electroanal. Chem. 259 (1989) 81-90

23. P.J. Kulesza, B. Grzybowska, M.A. Malik, M.T. Galkowski, Tungsten Oxides as active supports for highly dispersed paltinum microcenters: electrocatalytic reactivity toward reduction of hydrogen peroxide and oxygen, J. Electrochem. Soc. 144 (1997) 1911-1917

24. A.A. Karyakin, Electrochemical sensors, Biosensors and their Biomedical Applications, N.Y., 2008, p. 411

25. N.B. Li, J.H. Park, K. Park, S.J. Kwon, H. Shin, J. Kwak, Characterization and electrocatalytic properties of Prussian blue electrochemically deposited on nano-Au/PAMAM dendrimer-modified gold electrode, Biosensors and Bioelectronics 23 (2008) 1519–1526

26. R. Araminaitė, R. Garjonytė, A. Malinauskas, Electrocatalytic reduction of hydrogen peroxide at Prussian blue modified electrodes: a RDE study Journal of Solid State Electrochemistry, 14, 2010, 149 – 155

27. R. Mažeikienė, N. Gediminas; A. Malinauskas, Electrocatalytic reduction of hydrogen peroxide at Prussian blue modified electrode: An in situ Raman spectroelectrochemical study, Journal of Electroanalytical Chemistry, 660, 2011, 140-146

28. L. Cao, Y. Liu, B. Zhang, L. Lu, In situ Controllable Growth of Prussian Blue Nanocubes on Reduced Graphene Oxide: Facile Synthesis and Their Application as Enhanced Nanoelectrocatalyst for H2O2 Reduction, ACS Applied Materials & Interfaces 2, 2010, 2339-2346

29. W. J. Albery, M.L. Hitchman, Ring-disc Electrodes, Clarendon Press, Oxford, (1971)

30. A.J. Bard, L.R. Faulkner, Electrochemical Methods, John Wiley & Sons, Inc., (2001)

31. Z. Galus, Fundamentals of Electrochemical Analysis (2nd ed) Wiley, NY, (1994)

32. V. Stamenkovic, T.J. Schmidt, P.N. Ross, N.M. Markovic, Surface segregation effects in electrocatalysis: kinetics of oxygen reduction reaction on polycrystalline Pt3Ni alloy surfaces, J. Electroanal. Chem. 554-555 (2003) 191-199

33. M. Bron, p. Bogdanov, S. Fiechter, I. Dorbandt, M. Hilgendorff, H. Schleburg, H. Tributsch, Influence of selenium on the catalytic properties of ruthenium-based cluster catalysts for oxygen reduction, J. Electroanal. Chem. 500 (2001) 510-517

34. L. Stobinski, B. Lesiak, A. Malolepszy, M. Mazurkiewicz, B. Mierzwa, J. Zemek, P. Jiricek, I. Bieloshapk, Graphene oxide and reduced graphene oxide studied by the XRD, TEMand electron spectroscopy methods, Journal of Electron Spectroscopy and Related Phenomena 195 (2014) 145–154

35. R. Chen, Q. Zhang, Y. Gu, L. Tang, C. Li, Z. Zhang, One-pot green synthesis of Prussian blue nanocubes decorated reduced graphene oxide using mushroom extract for efficient 4-nitrophenol reduction Analytica Chimica Acta 853 (2015) 579–587

36. M. Watanabe, H. Igarashi, K. Yosioka, An experimental prediction of the preparation condition of Nafion-coated catalyst layers for PEFCs, Electrochim Acta 40 (1995) 329-334

37. M. Chojak, A. Kolary-Zurowska, R. Wlodarczyk, K. Miecznikowski, K. Karnicka, B. Palys, R. Marassi, P.J. Kulesza, Modification of Pt nanoparticles with polyoxometallate monolayers: Competition between activation and blocking of reactive sites for the electrocatalytic oxygen reduction, Electrochim. Acta 52 (2007) 5574-5581

38. P.J. Kulesza, K. Miecznikowski, B. Baranowska, M. Skutnik, S. Fiechter, P. Bogdanoff, I. Dorbandt, Tungsten oxide as Active Matrix for Dispersed Carbon-Supported RuSex Nanoparticles: Enhancement of the Electrocatalytic Oxygen Reduction, Electrochem. Comm. 8 (2006) 904-908

39. S. K. Zecevic, J.S. Wainright, M.H. Litt, S.Lj. Gojkovic, R.F. Savinell, Kinetics of $O_2$ Reduction on a Pt Electrode Covered with a Thin Film of Solid Polymer Electrolyte , J. Electrochem. Soc. 144 (1997) 2973-2982

40. N. Alonso-Vante, P. Bogdanoff, H. Tributsch, On the Origin of the Selectivity of Oxygen Reduction of Ruthenium-Containing Electrocatalysts in Methanol-Containing Electrolyte, J. Catal. 190 (2000) 240-246

41. Y. Li, Y. Zhao, H. Cheng, Y. Hu, G. Shi, L. Dai, L. Qu, Nitrogen-Doped Graphene Quantum Dots with Oxygen-Rich Functional Groups, J. Am. Chem. Soc. 134 (2012) 15-18.

42. D.A. Slanac, W.G. Hardin, K.P. Johnston, K.J. Stevenson, Atomic Ensemble and Electronic Effects in Ag-Rich AgPd Nanoalloy Catalysts for Oxygen Reduction in Alkaline Media, J. Am. Chem. Soc. 134 (2012) 9812-9815.

**Figure captions**

**Fig.1.** X-ray diffraction patterns for the Pt/rGO (a), rGO_PB (b), Pt/rGO_PB (c) electrocatalysts.

**Fig.2.** Typical HR-TEM images of rGO_PB (a), Pt/rGO (b), Pt/rGO_PB (c) nanoparticle.

**Fig.3.** Cyclic voltammetric characterization of (a) Pt/rGO nanoparticles and (b) Pt/rGO_PB nanoparticles (deposited on glassy carbon) recorded in deaerated 0.5 mol dm$^{-3}$ H$_2$SO$_4$ electrolyte. Scan rate 50 mV s$^{-1}$.

**Fig.4.** RDE voltammetric responses of PB-modified rGO supported Pt catalyst nanoparticles recorded at 10 mV s$^{-1}$ scan rate in oxygen saturated 0.5 mol dm$^{-3}$ H$_2$SO$_4$ at different rotation rates (from 400 to 2500 rpm).

**Fig.5.** Koutecky-Levich reciprocal plots (prepared using the data of Fig. 4) for the electroreduction of oxygen (at 0. 7 V) at catalytic layers (deposited on glassy carbon disks) containing PB-modified rGO supported Pt nanoparticles (red) and Pt/rGO (blue). Loadings of Pt: 10 µg cm$^{-2}$.

**Fig.6.** Current-potential RRDE curves for oxygen reduction recorded at (red curve) PB-modified rGO supported Pt nanoparticles, and (blue curve) rGO supported Pt nanoparticles. Scan rate 10 mV s$^{-1}$. Rotation rate, 1600 rpm.

**Fig.7.** Fraction of hydrogen peroxide (X$_{\%H2O2}$) produced during electroreduction of oxygen under the conditions of RRDE voltammetric experiment.

**Fig.8.** Numbers of transferred electrons (n) per oxygen molecule during electroreduction of oxygen under conditions of the RRDE voltammetric experiments as for Fig. 7.

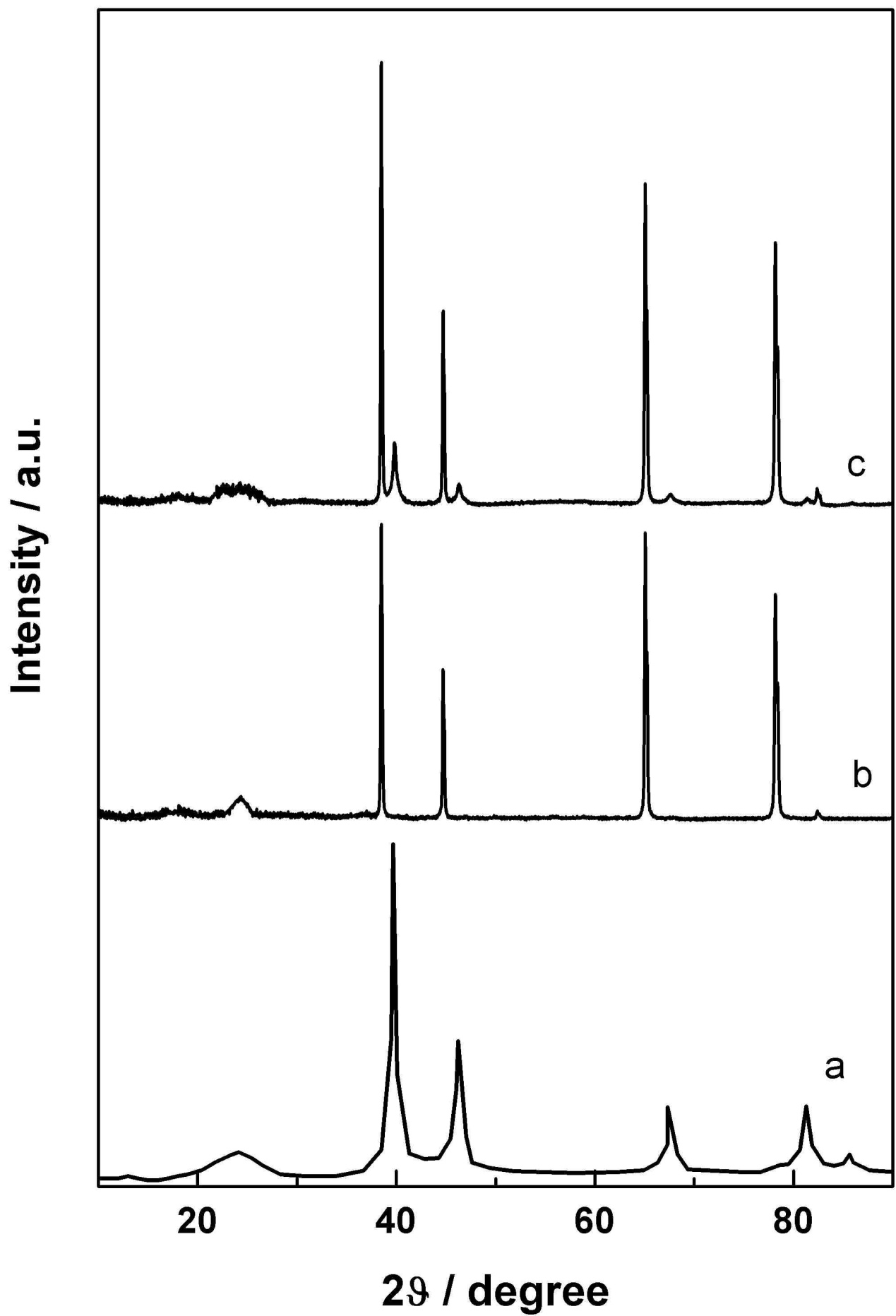

Fig.1

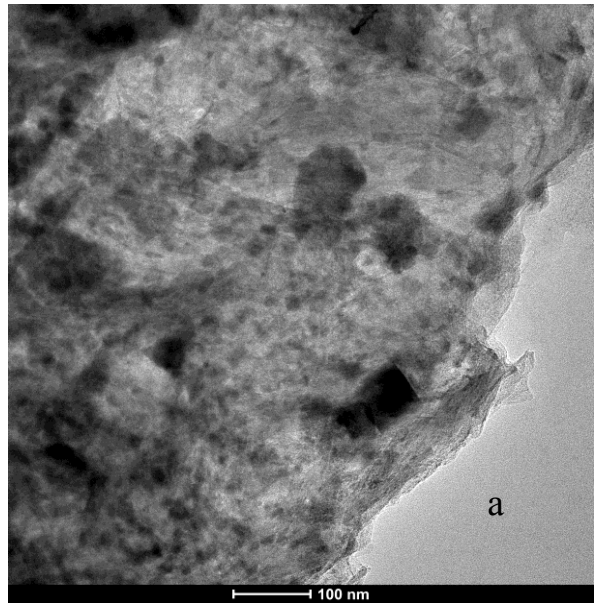
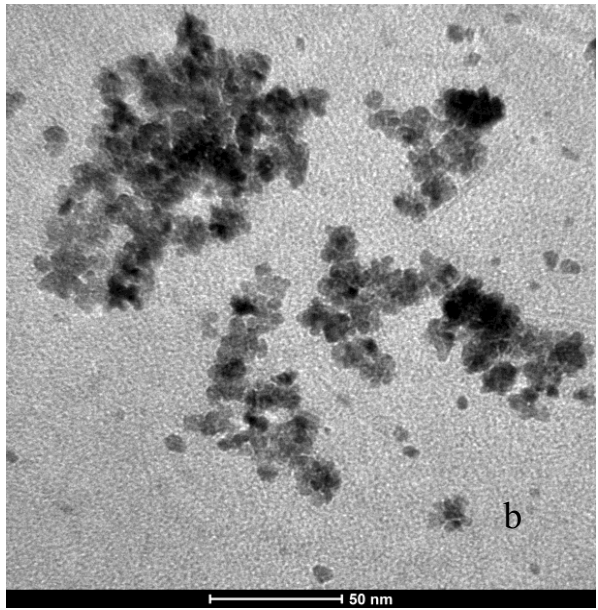
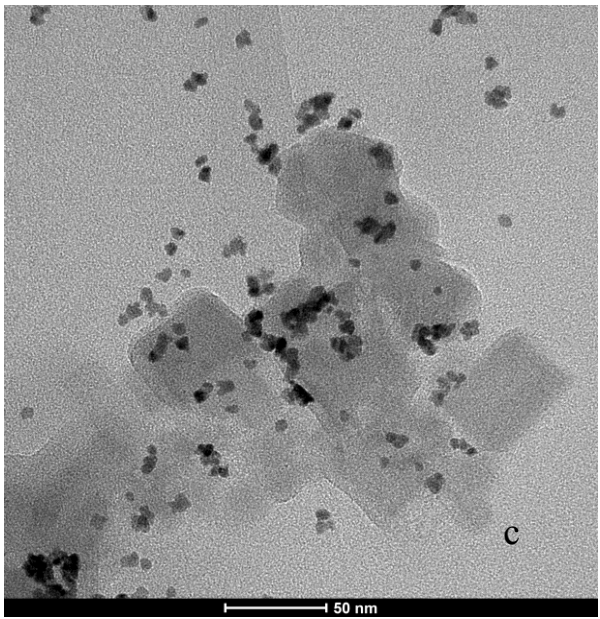

Fig.2

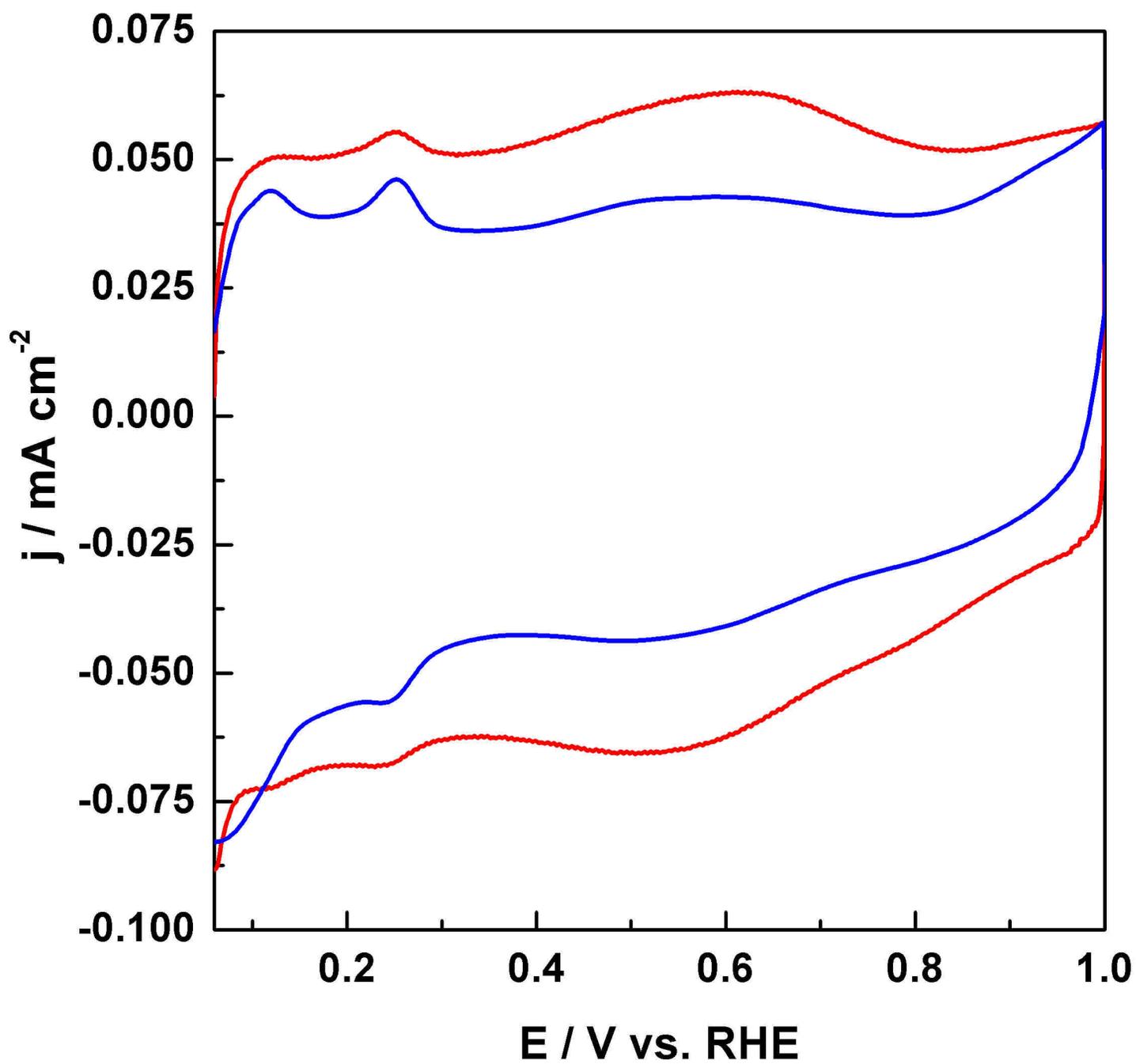

Fig.3

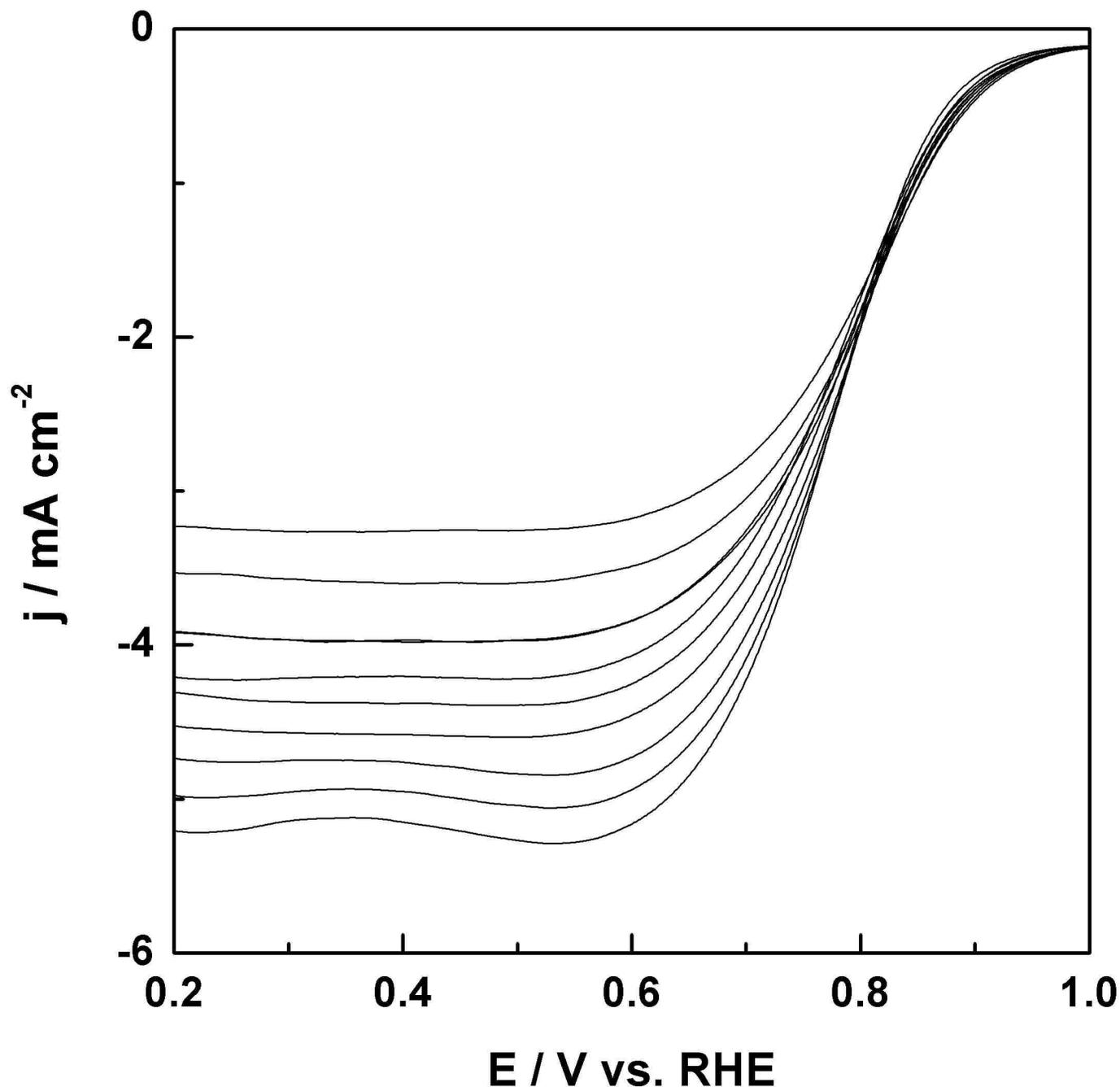

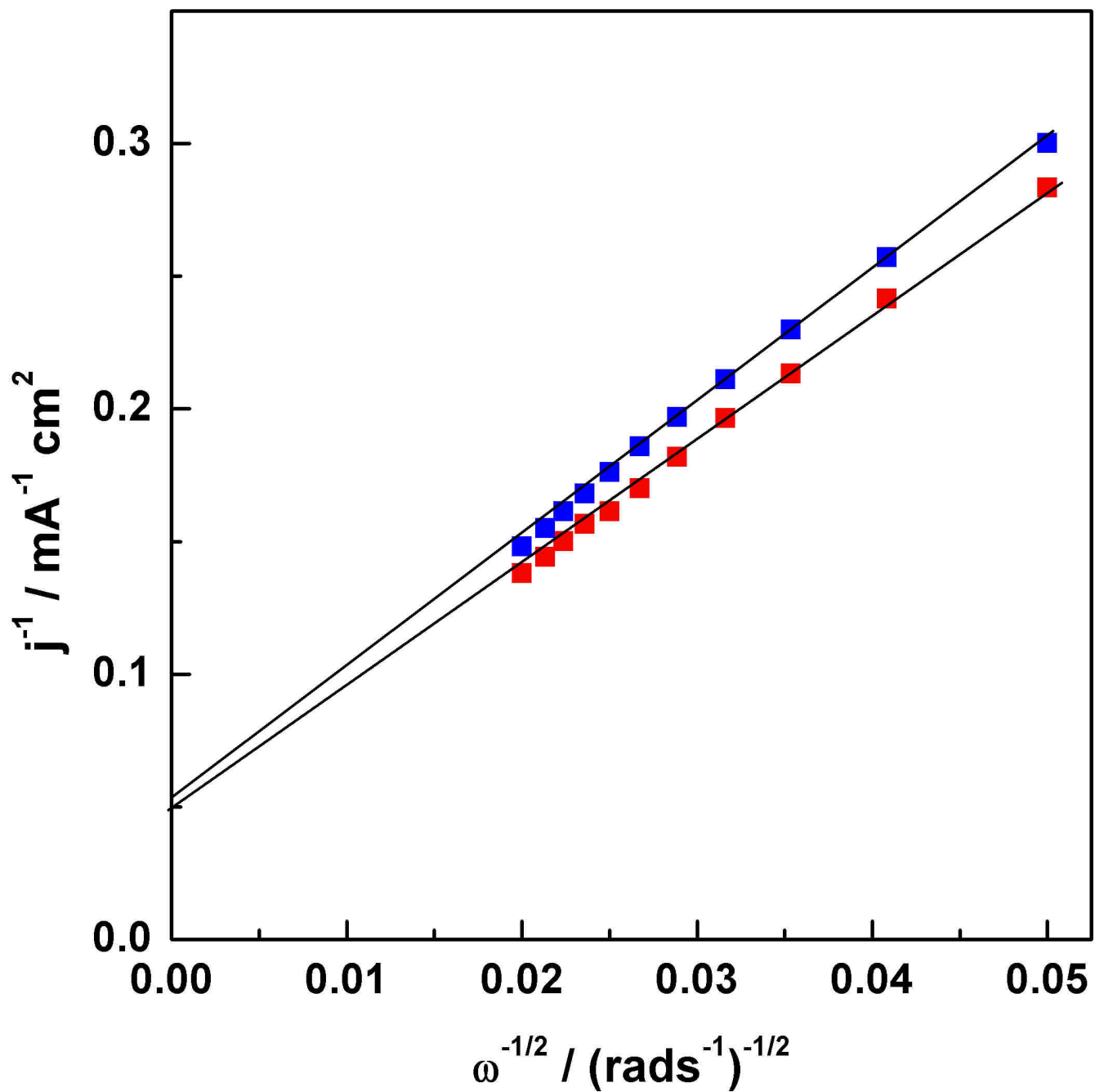

Fig.5

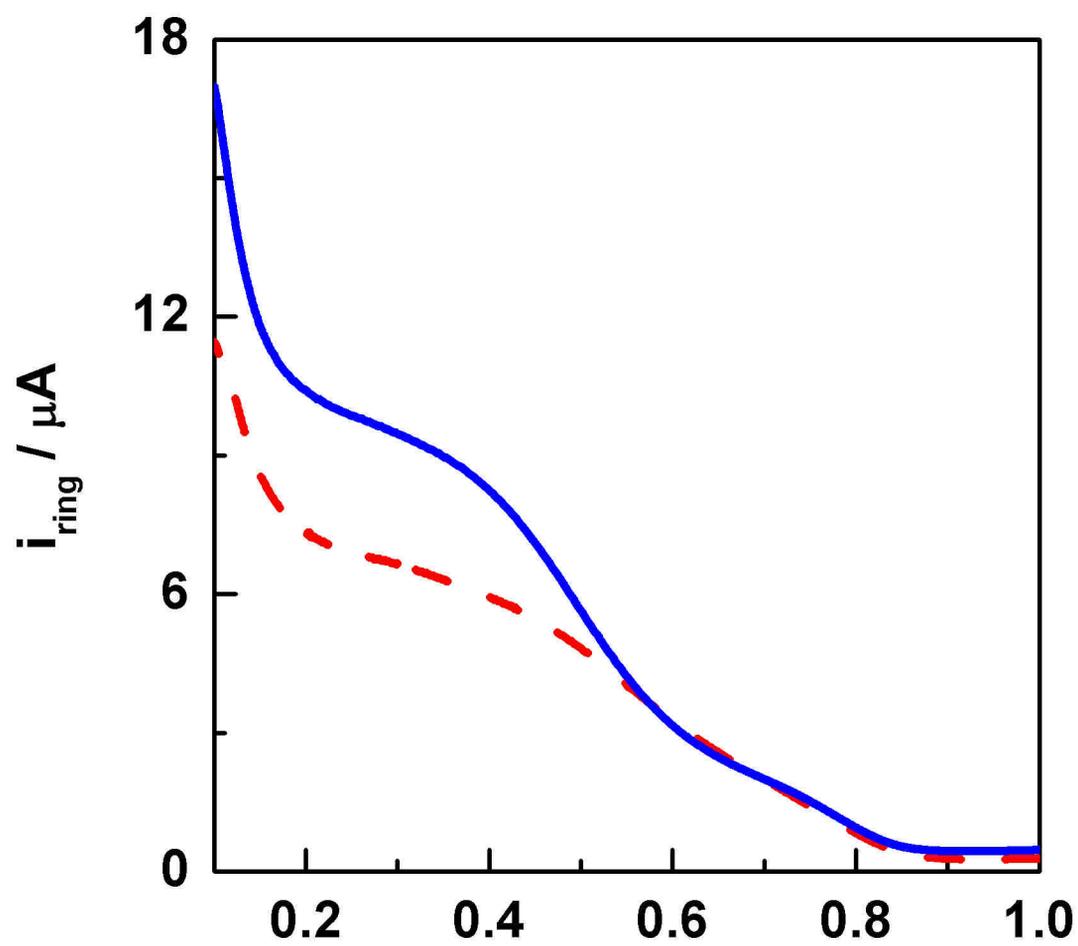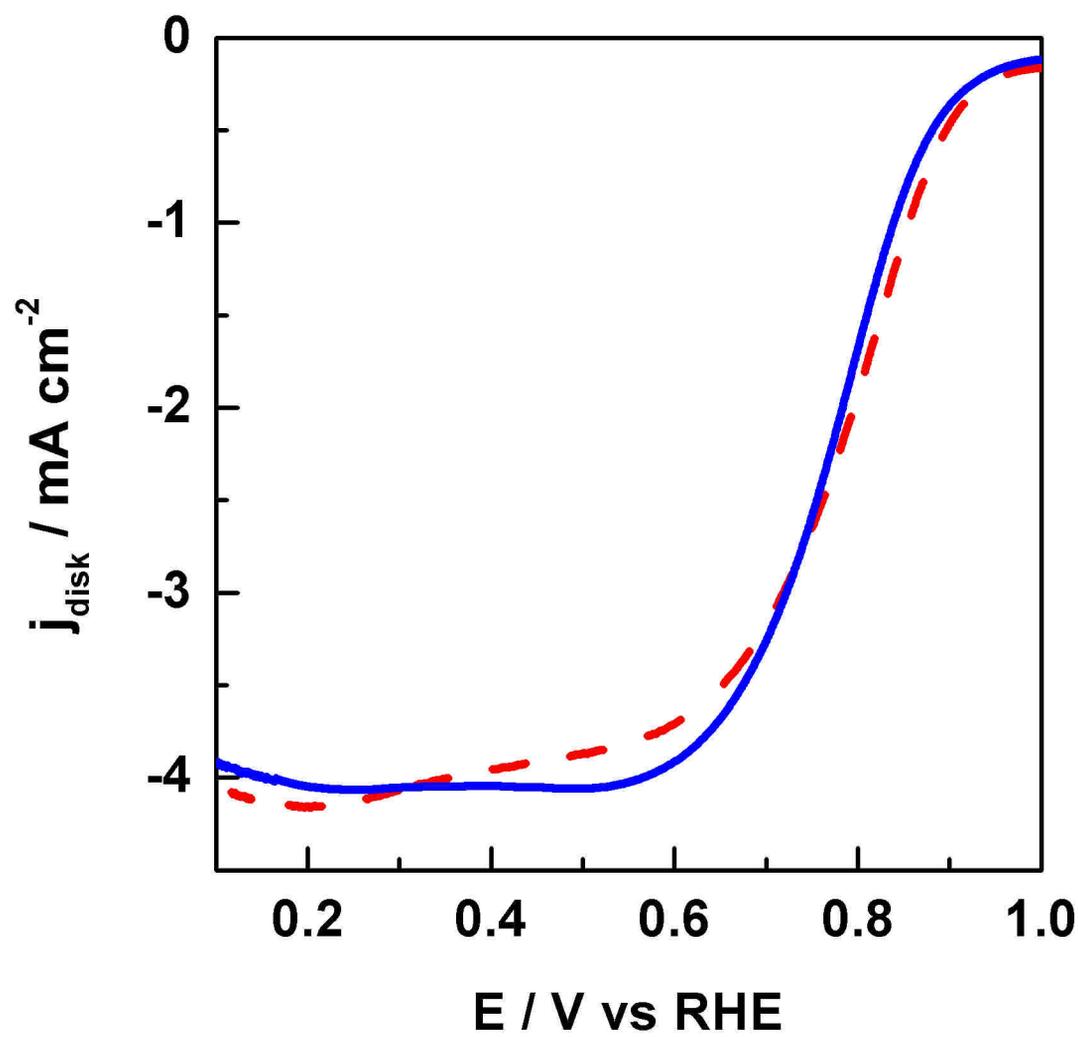

Fig.6

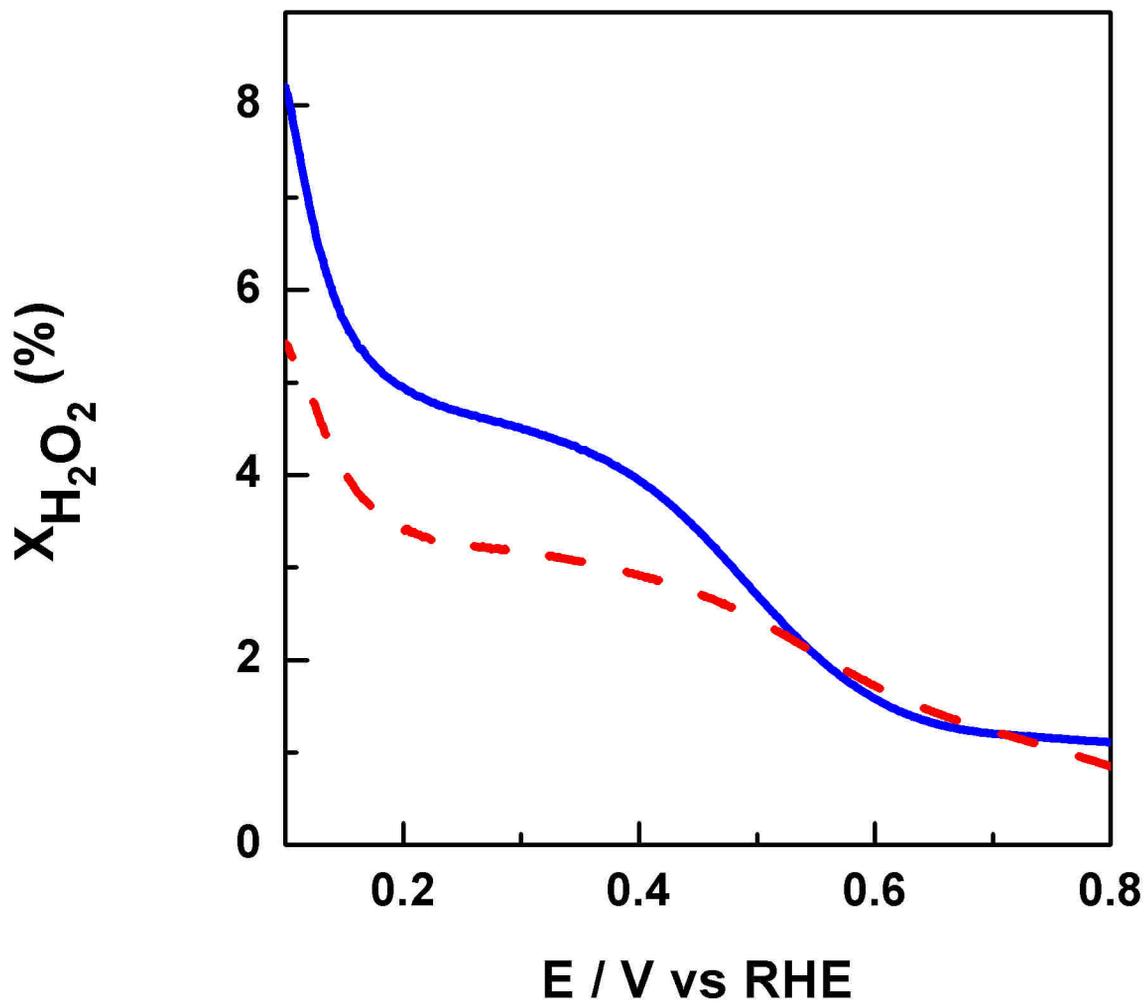

Fig.7

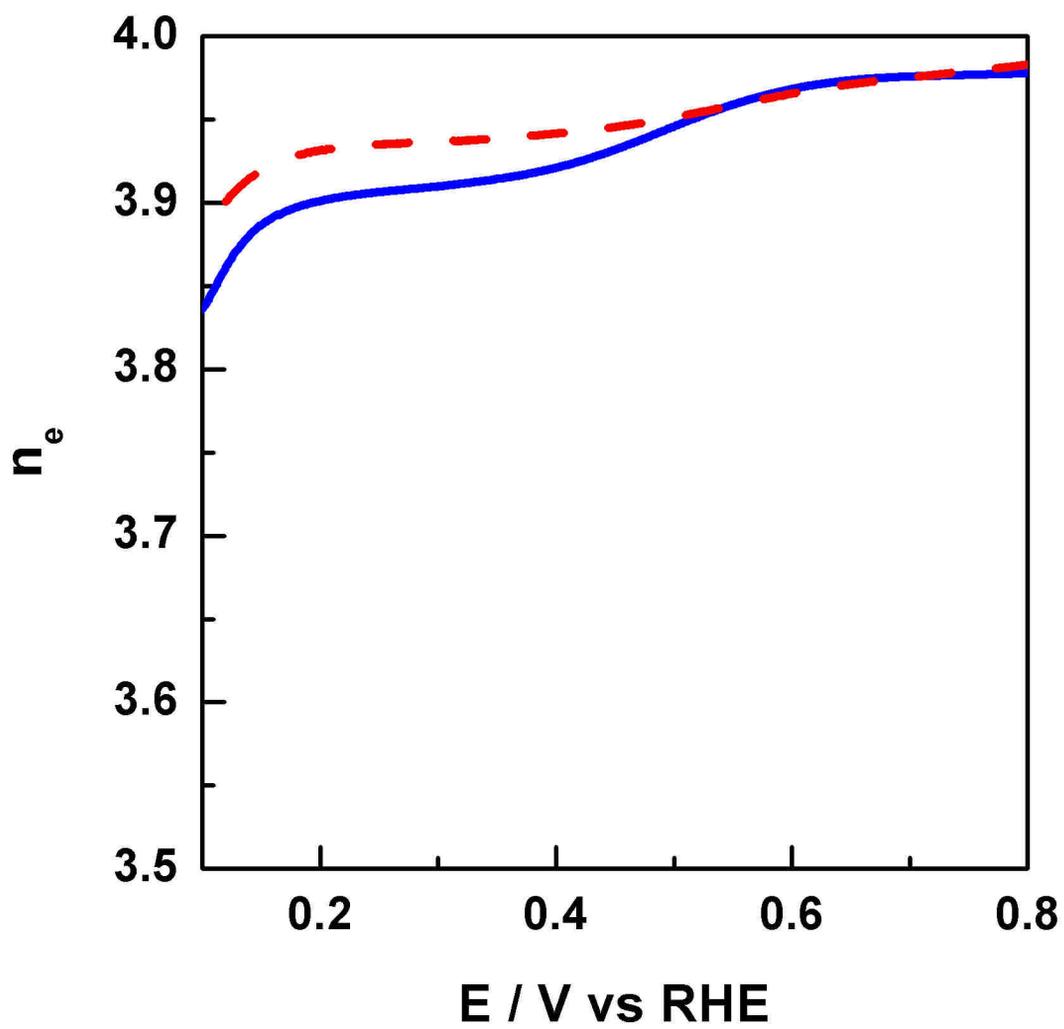

Fig.8